\documentclass[preprint,showpacs,preprintnumbers,amsmath,amssymb,floatfix]{revtex4}
\usepackage{graphicx}
\usepackage{dcolumn}
\usepackage{bm}
\usepackage{longtable}
\begin{document}

\title{Trends in  Ferromagnetism in Mn doped dilute III-V alloys from a density
functional perspective}
\author{Roby Cherian and Priya Mahadevan}
\affiliation{S.N. Bose National Centre for Basic Sciences, JD-Block, Sector III,
Salt Lake, Kolkata-700098, India} 
\author{Clas Persson}
\affiliation{Department of Materials Science and Engineering, Royal Institute 
of Technology, SE-10044, Stockholm, Sweden}
\date{\today}

\begin{abstract}
Mn doping in dilute III-V alloys has been examined as a route to enhance
ferromagnetic stability. Strong valence band bowing is expected at the
dilute limit, implying a strong modification of the ferromagnetic stability
upon alloying, with even an increase in some cases. Using first principle
electronic structure calculations we show that while codoping with a group 
V anion enhances the ferromagnetic stability in some cases when the effects
of relaxation of the lattice are not considered, strong impurity scattering
in the relaxed structure result in a reduction of the ferromagnetic stability.
\end{abstract}

\date{\today}
\maketitle

{\bf I. Introduction}

Dilute magnetic semiconductors have been intensively studied in recent 
times with the intention of replacing conventional electronic devices with 
those based on these materials \cite{review}. No real devices have been realized so far 
and the interest in this field has been based on concept devices. An 
essential aspect of dilute magnetic semiconductors based devices is the 
manipulation of the spin of the electron which one would like to do at room 
temperature. Hence the search is on for a room temperature ferromagnet. One 
material that has been intensively studied for this purpose is Mn doped GaAs
\cite{yu-nithin}. The ferromagnetic Curie temperature ($T_c$) are still 
far from desirable reaching a maximum of 250 K in specially 
designed superlattices \cite{superlattice}. Alternate materials 
such as transition metal atoms doped in II-VI semiconductors \cite{ando},
perovskite oxides \cite{sugata}, other oxides \cite{prl-nature}, III-V 
semiconductors and chalcopyrites \cite{medvedkin} have been 
synthesized and studied and the search is on for the most suitable material.
In this work we confine our attention to Mn doping in III-V semiconductors.
Theoretical calculations predict a 
higher ferromagnetic stabilization energy for Mn in GaN than in 
GaAs \cite{priya-apl}. However, the main conclusion from 
experiments is that the effective interaction between 
the Mn atoms is antiferromagnetic \cite{ploog1} or could be weakly 
ferromagnetic \cite{cilbert}. Strong ferromagnetism has been observed 
though the origin seems to be the presence of ferromagnetic clusters of Mn present at
large doping concentrations \cite{ploog2}.

In this work we examine an alternate strategy to obtaining high Curie 
temperatures by considering dilute alloys of III-V semiconductors. Strong
valence band bowing is expected in the dilute limit in some cases, and we 
want to use that to modify the ferromagnetic stability. These ideas 
are not new to the current work and have been proposed earlier in the 
literature \cite{nmat}. However there is no detailed theoretical work which has 
examined the electronic structure, modified interaction strengths 
and the consequent implications on the $T_c$. The basic idea that 
one aims to exploit here is to use semiconductors with band edges 
energetically closer to the Mn 3$d$ levels. This increases the 
hydridization between the host anion $p$ states and the Mn $d$ 
states. This however results in a deep acceptor level in the band gap. 
Thus the ferromagnetism-mediating holes are more localized. 
A system that has been proposed as a strong candidate is a 
dilute alloy of GaAs with GaP. The hope is that the itinerancy 
of the carriers is retained while the $p$-$d$ exchange is enhanced 
because of the shorter Mn-anion bond lengths. LDA+U/coherent potential 
approximation (CPA) based 
calculations \cite{masek} have looked at 25\%, 50\%, 75\% alloys of GaAs and GaP 
and found a modest increase in $T_c$. Experiments 
\cite{stone,stone-prb} have looked at the dilute limit of 
$Ga_{1-x}Mn_{x}As_{1-y}P_{y}$ and $Ga_{1-x}Mn_{x}As_{1-y}N_{y}$ 
with y $\sim$ 0.01-0.04. There is a strong decrease in $T_c$ 
with increasing y even at this dilute limit. 
As the dilute limit has not been studied theoretically we 
examine whether the experimental trends may be captured within our 
calculations and if not, then the cause is due to extraneous factors 
not included in the present approach. We are able to 
capture the reduction of $T_c$ in GaAs alloyed with GaN as well as 
the case of GaAs alloyed with GaP. 

{\bf II. Methodology}

In order to address these issues we have considered 216 atom supercells 
of GaAs, GaP and GaN. One Mn was placed at the origin while the second 
was placed at the fourth FCC neighbor position as earlier work has shown 
that the ferromagnetic stability is strongest for Mn atoms at these 
positions \cite{priya-prl}. 
We consider dilute III-V alloys as the host semiconductor into which Mn is 
doped. In order to form the dilute III-V alloys a group V anion impurity 
is codoped at different distances on the line perpendicular 
to the line joining the two Mn sites. For GaAs alloyed with GaP we 
included higher dopant concentrations also. 
The optimized lattice constants for the GaAs, GaP and GaN are 5.72, 5.49 
and 4.52 $\AA$ respectively. The lattice constant of the supercell is 
set according to Vegard's law for the III-V alloy. Full optimization 
of the internal positions is carried out within the first principle 
electronic structure calculations using a plane wave pseudopotential 
implementation of the density functional theory \cite{vasp}. 
A plane wave cutoff of 
400 eV was used for the basis set. The electronic structure was solved 
considering the generalised gradient approximation \cite{ggapw91} for the exchange 
at gamma point alone using projected augmented wave (PAW) 
potentials \cite{paw}.

{\bf III. Results and Discussion}

We first consider the case of the As doped into GaN. GaN we know has a large 
band offset with GaAs with the former having a deeper valence band maximum. 
Mn when doped into GaN has been shown to introduce states into the band gap 
with significant Mn character. However, in contrast, Mn when doped into 
GaAs is found to introduce states with weak Mn character in the band gap. 
Although in both cases the formal oxidation state of the Mn is 3+, in the 
former case the configuration is $d^{4}$ while in the latter case 
is \textbraceleft$d^{5}+$ hole\textbraceright \cite{priya-prb}. 
Since at this dilute limit, alloys show significant 
band bowing effects, we investigate Mn doping in this limit to probe 
modifications in the ferromagnetic stability. In Fig.1 we have plotted 
the Mn $d$ partial density of states. The N $p$ partial density of 
states of a N atom which is nearest neighbor of the Mn as well the 
As $p$ partial density of states have been shown. The valence band 
maximum seems to comprise of primarily As $p$ states with some N $p$ 
admixture. However there is hardly any As $p$ character at the 
Fermi level. We can understand this effect within a simple model 
that was proposed earlier to explain the electronic structure of transition-metal
(TM) impurities in semiconductors \cite{priya-prb}. The 
dominant interaction seems to be between the transition-metal 
impurity and its nearest neighbors. Atoms farther 
away from the TM impurity interact to a much lesser extent. 

The next question we ask was how is the ferromagnetic stability 
modified. Should the presence of the As impurity affect the ferromagnetic 
stability? As the As levels are between the Mn $d$ and the N $p$ levels
one would expect a modification in the ferromagnetic stability, possibly 
a value between the two end limits of Mn in GaN and GaAs. In this dilute 
alloy limit the valence band maximum is intermediate between that 
for GaAs and GaN. The ferromagnetic stability as well as the corresponding 
mean field estimate of $T_{c}$ are given in Table. I for  
various distances of the As impurity from either Mn atom. In the absence 
of As impurity, the ferromagnetic stability is $\sim$ 84 meV and 
at the other limit of Mn in GaAs it is $\sim$ 164 meV. This drastically 
drops to $\sim$ 21 meV in the presence of an As impurity at 3.76 $\AA$. 
As the As impurity is moved farther away
the ferromagnetic stability is partly regained though it still 
remains less than the value in the absence of As impurity. This is 
contrary to our expectations and indicates that one must consider 
other factors such as alloy scattering in addition to the 
modified energy denominator for the interaction.

Examining the opposite limit of GaAs into which the N is doped we find a 
similar trend in the ferromagnetic stabilization energy. In GaAs, Mn 
doping gives rise to a ferromagnetic stabilization energy of 164 meV 
(Table. II). This drastically drops to 100 meV (Table. II) for a 
N impurity introduced at 4.73 $\AA$. Ferromagnetic stability corresponding 
to the unalloyed limit is partially regained for the N atoms 
farther away from Mn as shown in Table. II. 

Recent experimental work \cite{stone} on dilute magnetic semiconductors have focussed 
on transition metal impurities in dilute alloys with an aim of  
understanding the mechanism of the ferromagnetism better. If the 
hole introduced by Mn doping is a valence band hole the belief 
is that it should be weakly perturbed by alloying effects at the dilute 
limit. However a hole which resides in an impurity band is expected to be strongly 
affected by alloying effects. We artificially tune the hole introduced by the 
Mn doping with a introduction of $U$ on the Mn $d$ states in $(Ga,Mn)As_{0.99}N_{0.01}$. 
With a $U$ of 4 eV on the Mn $d$ states, the hole moves towards a valence band 
hole. Hence alloy scattering effects are expected to be weaker. Indeed our 
calculated ferromagnetic stability results effect this. The ferromagnetic (FM)
stability results reflect this. The FM stability changes by about 64 meV 
in the absence of $U$ to 15 meV in the presence of $U$.
Stone et al. \cite{stone} have examined 
P doping in GaMnAs. An insulator to metal transition is observed as a 
function of doping. Usually high effective mass have to be assumed 
for the carriers to explain the experimental 
observations. This reinforces the idea of the carriers residing 
in an impurity band.

While Mn in GaN has usually been accepted as a system in which the 
hole introduced by Mn doping resides in an impurity band, the case 
of Mn in GaAs is heavily debated with supporters on both sides. 
In order to quantify our observations further we examined Mn doping 
in dilute alloys formed by P introduction in GaAs. If our earlier 
results of the N doping in GaAs established the validity of the 
impurity band model, then we should see strong effects on the 
ferromagnetic stability here also. The results as a function 
of the impurity distance are given in Table. III. The perturbation 
seems to be very weak and consequently the variations in the 
ferromagnetic stabilization energy from the unperturbed case are small. 
Hence modifications of ferromagnetic stability in alloyed systems 
is not proof enough for the impurity band picture. The deviation 
between our results and experimental results of $GaMnAs_{1-y}P_{y}$ 
could be due to various reasons. One cause could be that the alloying 
concentration are large. Indeed when we introduced two P impurities, such 
that each P atom is the nearest neighbor to one of the Mn atom and also 
these P atoms connects the Mn atoms via a Ga atom, we found a 
reduction in the ferromagentic stability from 164 meV in the unalloyed 
limit to 110 meV. Thus impurity scattering is responsible for the 
reduction in ferromagnetic stability.

Closer analysis revealed that there are two parts which 
need to be considered when an impurity atom is introduced. The first part 
is that associated with the modified electronic interaction strengths 
associated with the impurity and the second associated with the 
strain in the host lattice. Indeed we cannot decouple the two effects 
completely but the effects of these perturbations may be discussed 
within calculations performed under some constraints.

In Table. IV we consider the case of N codoped into GaAs at a distance 
of 4.73 $\AA$. The ferromagnetic stability is evaluated in the 
unrelaxed case where the calculation is performed assuming that 
N merely replaces an As atom. 
This calculation would capture the effects of modified interaction 
strengths as a result of N doping. Ferromagnetic stability 
of the Mn pairs is increased from 164 meV in the absence of N impurity 
to 203 meV. Examining the density of states corresponding to the unrelaxed 
limit, we find that N $p$ states are energetically closer to the Mn $d$ 
states and hence the increased interaction between the two could explain the
increased ferromagnetic stability.
This is reflected in the inset of Fig. 2(c) which shows the 
N $p$ character near Fermi energy.
Ga-N bonds are much smaller than the Ga-As bonds. 
Allowing the atoms to optimize their internal positions by total energy 
minimization We find 
that the Ga-N bonds are $\sim$ 2.08 $\AA$ long while the 
Ga-As bonds close to the N atom are 2.53 $\AA$ long. Those 
far away from the N impurity are 2.48 $\AA$ long. Examining 
the ferromagnetic stability in such a configuration, 
we find that it is drastically 
reduced to 100 meV. Examining the density of states (Fig. 2) 
we find that there is hardly any change in the As $p$ density of 
states plotted for the nearest neighbor As atom as well as the 
Mn $d$ partial density of states with and without relaxation. 
N $p$ density of states show significant changes with relaxation 
with the N $p$ states moving from -1 eV to -3.5 eV below the 
Fermi level. In the inset of Fig. 2 (c) we magnify the near Fermi 
energy region of the N $p$ density of states and see that 
there is significant N $p$ character at the Fermi energy in the 
unrelaxed case which is reduced upon relaxation. Since movement 
of N $p$ levels should imply reduced interaction with Mn, the 
main effect of ferromagnetic stability reduction has to be 
the lattice strain.
Considering the case of P in GaAs doped with 
Mn (Fig. 3), we find that the effects of relaxation are very 
weak as a result of which ferromagnetic stability is hardly affected 
(Table. III). Thus we are able to elucidate the microscopic 
origin of the reduction of $T_{c}$ when we alloy III-V semiconductors 
with another group V anion. Although our calculations provide us
with qualitative trends, an exact numerical expression 
for the $T_{c}$ as a function of concentration in the
dilute limit is difficult.  This is because the 
ferromagnetic stability is a strong function of
the disorder position as well as of concentration. Strong disorder
potentials caused by clustering of the alloying atoms results in a
stronger decrease in $T_{c}$ than when one has a random
distribution.

{\bf IV. Conclusion}

We have examined Mn doping in the dilute alloy limit where 
strong valence band bowing is observed, as a possible route 
to enhance the ferromagnetic transition temperature. This has been 
quantified in our calculation as the ferromagnetic stabilization 
energy for a pair of Mn atoms occupying lattice sites for which 
ferromagnetic stability has been observed to be strong. Contrary 
to expectations of enhanced ferromagnetic stability on alloying 
we find a reduction in the case of $Ga_{1-x}Mn_{x}As_{1-y}N_{y}$ and 
$Ga_{1-x}Mn_{x}As_{1-y}P_{y}$ where y $\sim$ 0.01. Mn doping in 
$Ga_{1-x}Mn_{x}As_{1-y}P_{y}$ shows very small changes in the 
ferromagnetic stability for 1\% anion doping but for 
larger percentage of doping shows a reduction in $T_{c}$.
The origin of reduced ferromagnetic 
stability is traced to the strong strain effects that accompany 
the introduction of the anion impurity. This strongly scatters 
the electron comprising the valence band maximum and therefore 
modifies the ferromagnetic stability. 

PM thanks the Indo French centre for financial support. 
CP thanks Swedish Energy Agency and 
Swedish Foundation for International Cooperation in Research and Higher
Education (STINT).

\renewcommand

\newpage
\begin{figure}
\includegraphics[width=5.5in,angle=270]{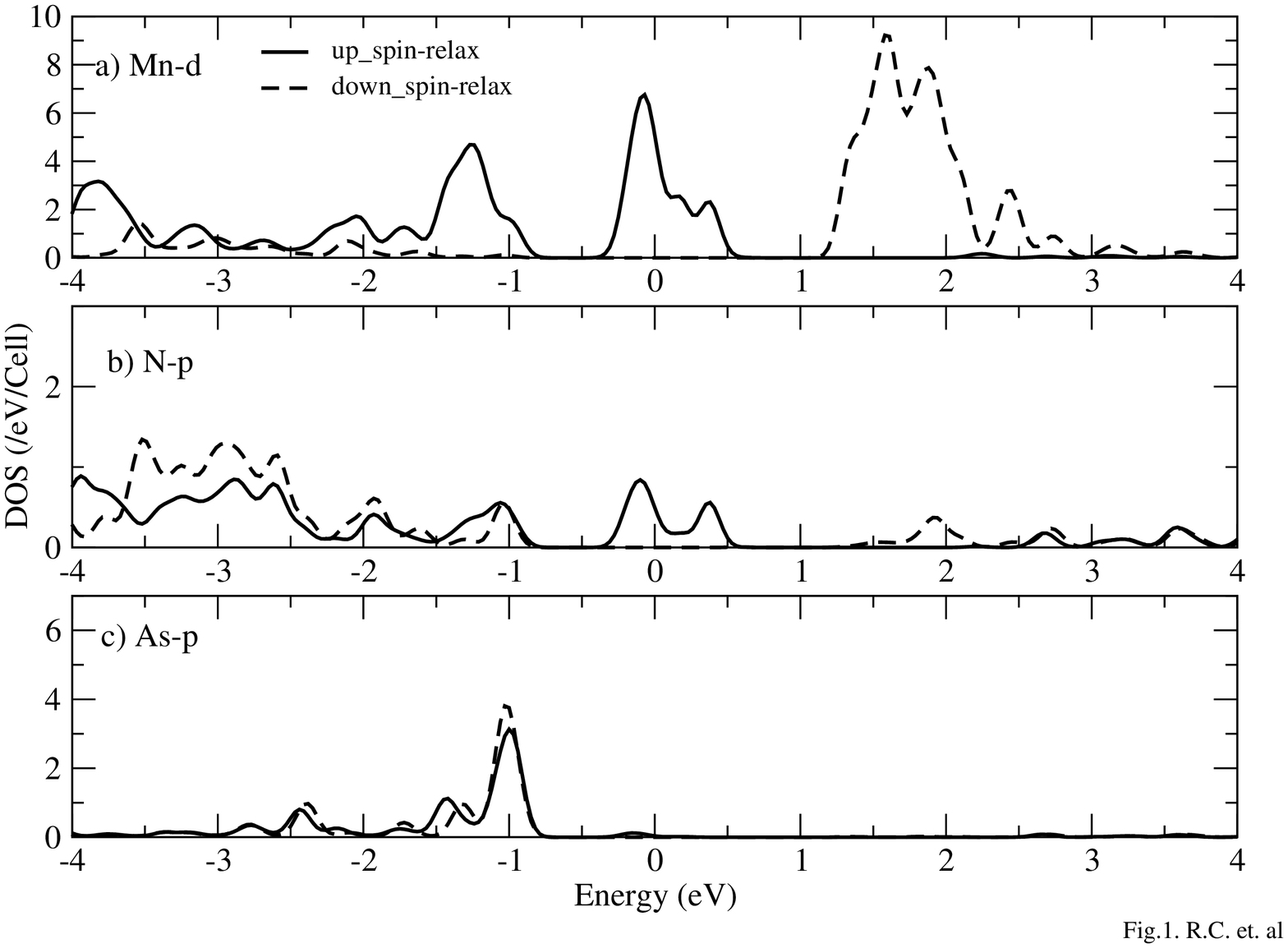}
\caption{(a) The up (solid line) and down (dashed line) spin 
projected density of states for Mn $d$, (b) N $p$, 
which is the nearest neigbor of the Mn, and (c) for 
the As $p$ calculated for (Ga,Mn)N codoped with As.
The zero of energy corresponds to the Fermi energy.}
\end{figure}

\newpage
\begin{figure}
\includegraphics[width=5.5in,angle=270]{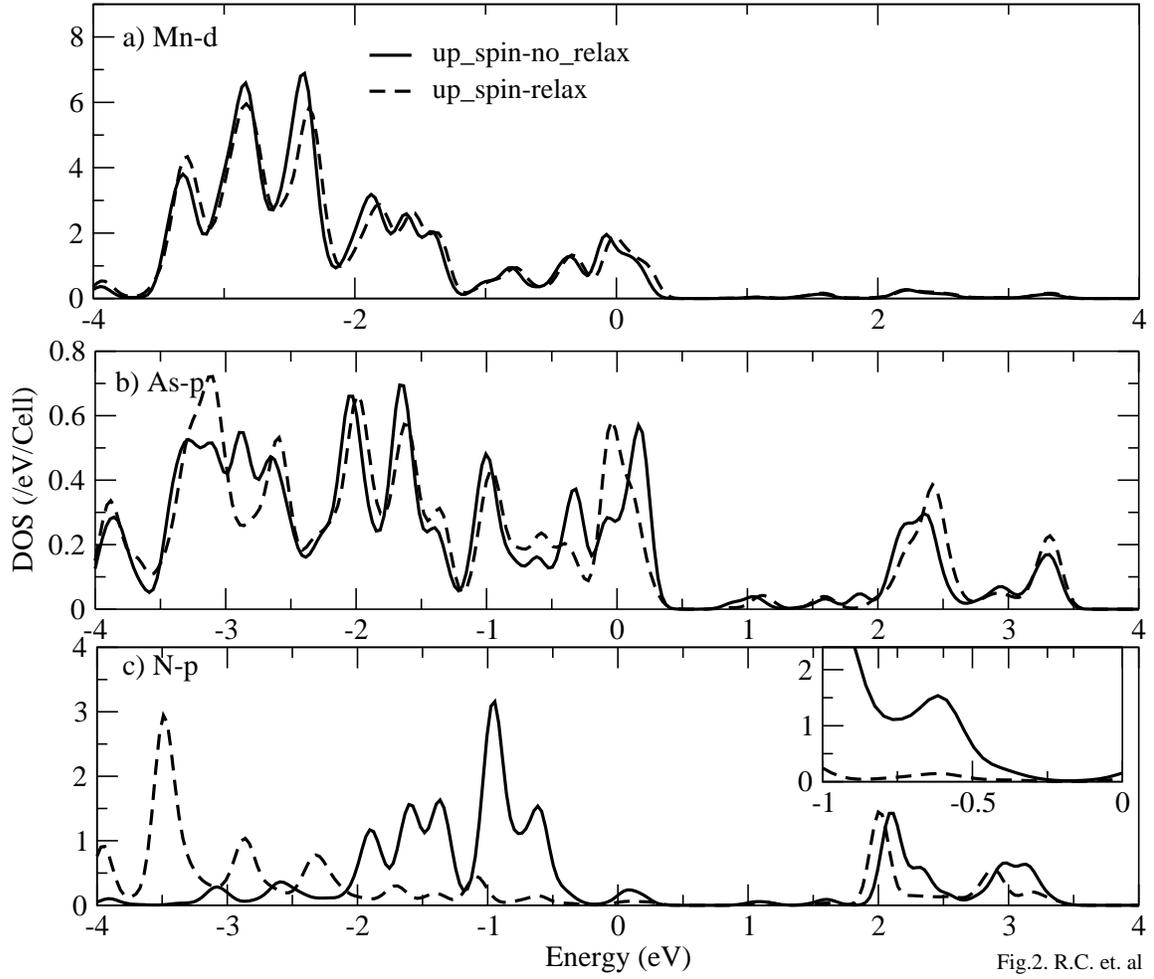}
\caption{(a) The spin up projected density of states for 
Mn $d$, (b) As $p$, which is the nearest neigbor of the Mn, and (c)
for the N $p$ calculated for (Ga,Mn)As codoped with N.
Inset of (c) shows a magnified 
view of the N $p$ density of states near the Fermi energy region. 
The solid lines corresponds to the unrelaxed case while the 
dashed lines corresponds to the relaxed case.
The zero of energy corresponds to the Fermi energy.}
\end{figure}

\newpage
\begin{figure}
\includegraphics[width=5.5in,angle=270]{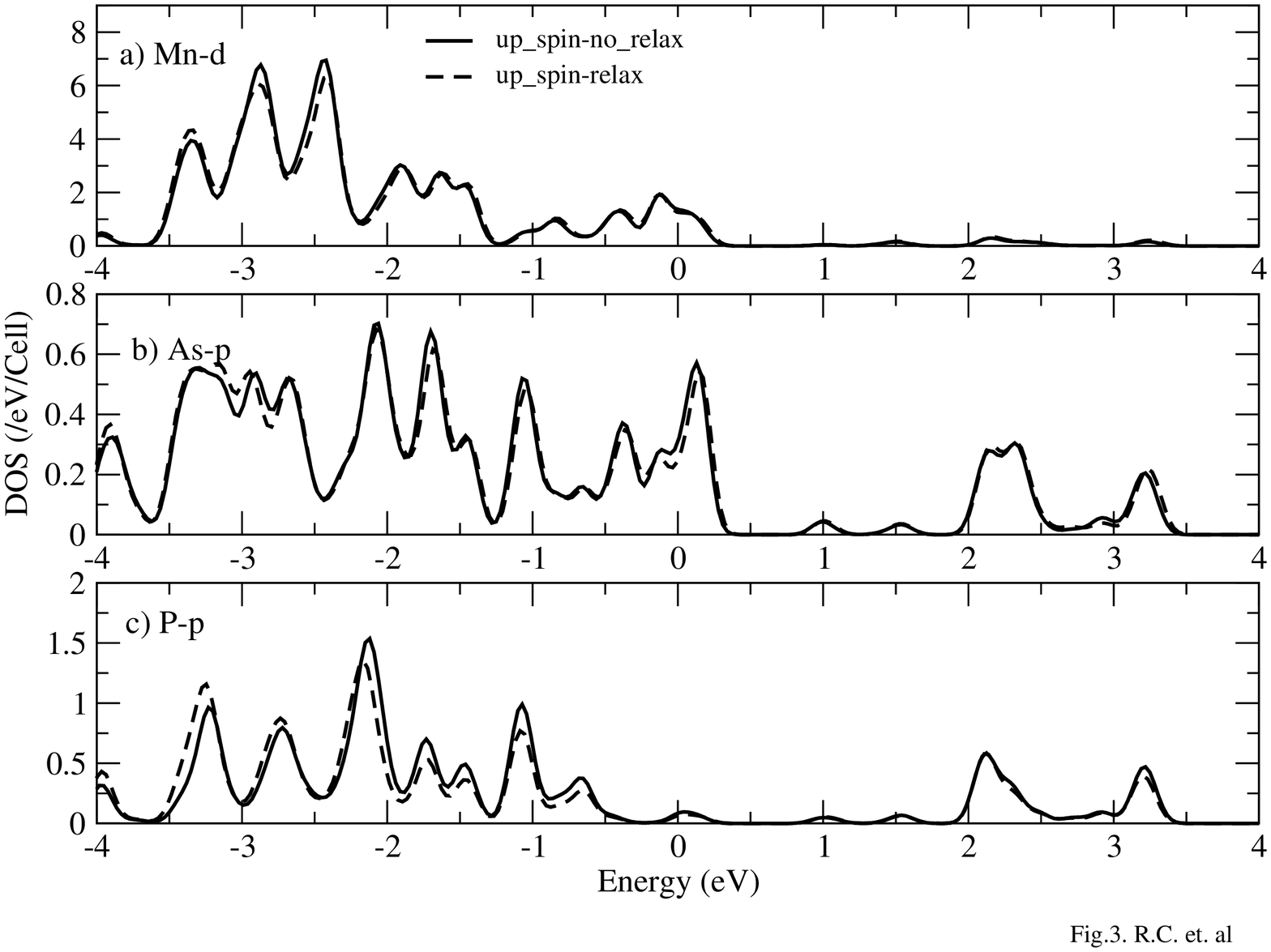}
\caption{(a) The spin up projected density of states for 
Mn $d$, (b) As $p$, which is the nearest neigbor of the Mn, and (c)
for the P $p$ calculated for (Ga,Mn)As codoped with P.
The solid lines corresponds to the unrelaxed case while the 
dashed lines corresponds to the relaxed case.
The zero of energy corresponds to the Fermi energy.}
\end{figure}

\newpage

\begin{table}
\caption { Effect on the stability by As codoping in
(Ga,Mn)N host. The distance of the As atom from either Mn atoms in both the 
ferromagnetic (FM) and antiferromagnetic (AFM) relaxed configuration is also
given. }
\begin{tabular}{c|c|c|c|c}
\hline\hline
 \multicolumn{2}{c} {Distance ($\AA$)} \vline&  Ferromagnetic stability (meV) & $T_{c}$ (K)   \\
\cline{1-4}
FM & AFM & ($E_{FM}-E_{AFM}$) & \\
\hline
  - &  - & -83.8 & 324.2 \\
  3.72  & 3.71 &  -20.6 & 79.6 \\
  5.89 & 5.86 & -50.0 & 193.4 \\
  6.71 & 6.69 & -50.0 & 193.4  \\
  8.11 & 8.10 & -34.8 & 134.6\\
\hline
\end{tabular}
\end{table}

\begin{table}
\caption { Effect on the stability by N codoping in
(Ga,Mn)As host. The distance of the N atom from either Mn atoms in both the
ferromagnetic (FM) and antiferromagnetic (AFM) relaxed configuration is also
given. }
\begin{tabular}{c|c|c|c|c}
\hline\hline
 \multicolumn{2}{c} {Distance ($\AA$)} \vline&  Ferromagnetic stability (meV) & $T_{c}$ (K)  \\
\cline{1-4}
FM & AFM & ($E_{FM}-E_{AFM}$) & \\
\hline
  - & -  &-163.8 & 633.6\\
  4.73 & 4.75 &  -100.1 & 387.2 \\
  7.41 & 7.42 & -154.3 & 596.9\\
  8.44 & 8.44 & -161.7 & 625.5 \\
  10.18 & 10.19 & -147.5 & 570.6 \\
\hline
\end{tabular}
\end{table}

\begin{table}
\caption { Effect on the stability by P codoping in
(Ga,Mn)As host. The distance of the P atom from either Mn atoms in both the
ferromagnetic (FM) and antiferromagnetic (AFM) relaxed configuration is also
given. }
\begin{tabular}{c|c|c|c|c}
\hline\hline
 \multicolumn{2}{c} {Distance ($\AA$)} \vline&  Ferromagnetic stability (meV) & $T_{c}$ (K)  \\
\cline{1-4}
FM & AFM & ($E_{FM}-E_{AFM}$) & \\
\hline
  - & - & -163.8 & 633.6\\
  4.73 & 4.75  &  -154.3 & 596.9 \\
  7.42 & 7.44 & -162.1 & 627.0\\
  8.45 & 8.45  & -169.1 & 654.1\\
  10.21 & 10.21 & -172.7 & 668.0\\
\hline
\end{tabular}
\end{table}

\begin{table}
\caption{ Effects of relaxation on the stability in the case of (Ga,Mn)As 
host codoped with N.}
\begin{tabular}{c|c|c|c|c|c}
\hline\hline
   & \multicolumn{2}{c} {Energy (eV)}\vline &  Ferromagnetic stability (meV) & $T_{c}$ (K)  \\
\cline{2-5}
  & $E_{FM}$  &  $E_{AFM}$  &  ($E_{FM}-E_{AFM}$) & \\
\hline
Un-Relaxed & -907.192 & -906.989 & -203.2 & 786.0\\
Relaxed & -909.264 & 909.164 & -100.1 & 387.2 \\
\hline
\end{tabular}
\end{table}

\end{document}